\documentstyle[prl,aps,preprint,tighten,floats,epsf]{revtex}

\newbox\rotbox

\begin{document}
\draft
\setcounter{page}{0}
\def\footnoterule{\kern-3pt \hrule width\hsize \kern3pt}
%
\title{INTERFERENCE FRAGMENTATION FUNCTIONS AND VALENCE QUARK SPIN
DISTRIBUTIONS IN THE NUCLEON\thanks
{This work is supported in part by funds provided by the U.S.
Department of Energy (D.O.E.) under cooperative 
research agreement \#DF-FC02-94ER40818, and by the RIKEN BNL
Research Center.}}

\author{R.~L.~Jaffe$^{\rm a,b}$,\footnote{Email address: {\tt jaffe@mitlns.mit.edu}}
 Xuemin~Jin$^{\rm a}$,\footnote{Email address: {\tt jin@ctpa02.mit.edu}}
 and Jian~Tang$^{\rm a}$\footnote{Email address: {\tt jtang@mitlns.mit.edu}} }

\address{{~}\\$^{\rm a}$Center for Theoretical Physics\\ 
Laboratory for Nuclear Science \\ 
and Department of Physics \\
Massachusetts Institute of Technology \\
Cambridge, Massachusetts 02139 {~} \\ 
{~}\\$^{\rm b}$RIKEN BNL Research Center \\
Brookhaven National Laboratory, Upton, NY 11973\\
{~}}

\date{MIT-CTP-2690 ~~~~ hep-ph/9710561{~~~}October 1997}
\maketitle

\thispagestyle{empty}

\begin{abstract}

We explore further applications of the twist-two quark interference 
fragmentation functions introduced earlier. We show that 
semi-inclusive production of two pions in the current fragmentation region 
in deep inelastic scattering of a longitudinally polarized electron on 
a longitudinally polarized nucleon can provide a probe of the valence 
quark spin (or helicity difference) distribution in the nucleon. 

\end{abstract}

\vspace*{\fill}
\begin{center}
Submitted to: {\it Physical Review D}
\end{center}

Measurements of quark and gluon distributions within hadrons provide
us with valuable information about the nonperturbative nature of
the quarks and gluons inside the hadrons. In a recent Letter
\cite{jjt1}, we have studied semi-inclusive production of
two pions in the current fragmentation region in deep inelastic scattering
on a transversely polarized nucleon. This may provide a practical way to
measure the quark transversity distribution in the nucleon, which 
has proved difficult to access experimentally \cite{ralston79,artru93,%
jaffe91,collins93,ji94,collins94,jaffe96}.
In this paper, we extend our study to the case of a 
longitudinally polarized electron beam scattering off
a longitudinally polarized nucleon target.
We show that the interference 
between the $s$- and $p$-wave of the two-pion system around the 
$\rho$ can provide an asymmetry [Eq.~(\ref{asymmetry})] which is sensitive 
to the valence quark spin distribution in the 
nucleon. Note that the asymmetry would vanish by C-invariance
if the two pions  were in a charge conjugation eigenstate.  
Hence there is no effect in regions of the $\pi\pi$  mass dominated 
by a single resonance. Significant effects are possible, however, in 
the $\rho$ mass region where the $s$- and $p$-wave 
production channels are both active and provide exactly the charge 
conjugation mixing necessary.  

The asymmetry we obtain throws pions of one charge 
forward along the fragmentation axis relative to pions of the other charge.  
If one integrated over the other kinematic variables, whatever result 
persisted would appear as a difference between the $\pi^+$ and $\pi^-$ 
fragmentation functions correlated with the valence quark spin
distributions.  This effect in single pion fragmentation was proposed and 
studied some years ago by Frankfurt {\it et al.} \cite{frankfurt89} 
and Close and Milner \cite{close91}.  The asymmetry we describe is 
therefore 
one particular contribution to this more general effect, 
with the advantage that it can 
be characterized in terms of $\pi\pi$ phase shifts and two particle 
fragmentation functions that appear in other hard processes.


Consider the semi-inclusive deep inelastic scattering process:
$\vec e\vec N\rightarrow e'\pi^+\pi^- X$.  We define the 
kinematics as follows. The four-momenta of the initial and final electron are
$k = (E, {\vec k})$ and $k^\prime = (E^\prime, {\vec k^\prime})$, and
the nucleon's momentum is $P_\mu$. The momentum of the virtual photon
is $q=k-k^\prime$, and $Q^2 = -q^2=-4EE'\sin^2{\theta/2}$, where
$\theta$ is the electron scattering angle. 
 The standard variables in DIS, $x = Q^2/2P\cdot
q$ and $y = P\cdot q/P\cdot k$, are adopted. We work at low $\pi\pi$ 
invariant mass, where only the $s$- and $p$-waves are significant.
The $\sigma [(\pi\pi)^{I=0}_{l=0}]$ and $\rho [(\pi\pi)^{I=1}_{l=1}]$
resonances  are produced in the current fragmentation
region with momentum $P_h$ and momentum fraction $z = P_h\cdot
q/q^2$.\footnote{We recognize that the $\pi\pi$ s-wave is not resonant in the 
vicinity of the $\rho$ and our analysis does not depend on a resonance 
approximation.  For simplicity we
refer to the non-resonant $s$-wave as the ``$\sigma$''.}
The invariant squared mass of the two-pion system is $m^2 =
(k_++k_-)^2$, with $k_+$ and $k_-$ the four-momentum of $\pi^+$
and $\pi^-$, respectively. The decay polar angle in the rest frame of
the two-pion system is denoted by $\Theta$. Note that the azimuthal
angle $\phi$ of the two-pion system does not figure in present 
analysis and can be integrated out.

Following Ref.~\cite{jjt1}, we use a collinear 
approximation, {\it i.e.}, 
$\theta\approx 0$ for simplicity, and work only to the leading twist 
(the complete analysis will be published elsewhere\cite{JJT2}).  
Invoking the helicity density matrix formalism developed in 
Refs.~\cite{jaffe95,jaffe96}, we factor the process into various 
basic ingredients expressed as helicity density matrices:
\begin{eqnarray}
\left[{{d^5\sigma}\over{dx\, dy\, dz\, dm^2\, d\cos\Theta}}\right]_{H'H}&&
={\cal F}\,_{H'H}^{h_1h'_1}\left[{{d^2\sigma(e
q\rightarrow e' q')}
\over{dx\, dy}}\right]_{h'_1h_1}^{h_2h'_2}
\left[{{d^2\hat{\cal M}}\over{dz\,
dm^2}}\right]_{h'_2h_2}^{H_1H'_1}
\left[{{d{\cal D}}\over{d\cos\Theta}}\right]_{H'_1H_1}\ ,
\label{hme} 
\end{eqnarray}
where $h_i(h_i')$ and $H(H')$ are indices labeling the helicity states of
quark and nucleon, and $H_1(H'_1)$ labeling the helicity state of the resonance
($\sigma$, $\rho$). Physically, the four factors 
on the right-hand side of Eq.~(\ref{hme}) represent the $N\rightarrow q$ 
distribution function, the hard partonic $eq\rightarrow e'q'$ cross section,  
the $q \rightarrow (\sigma, \rho)$ fragmentation, and the decay 
$(\sigma, \rho)\rightarrow \pi\pi$, respectively. In order to incorporate 
the final state interaction, we have separated the 
$q\rightarrow \pi^+\pi^-$ fragmentation process into 
two steps. First, the quark fragments into the resonance ($\sigma$,
$\rho$),  then the resonance decays into two pions (see Figure 1 of 
Ref.~\cite{jjt1}).

The $s$-$p$ interference fragmentation function
describes the emission of a $\rho (\sigma)$ with helicity $H_1$ from a quark 
of helicity $h_2$, followed by absorption of $\sigma (\rho)$, with helicity 
$H'_1$ forming a quark of helicity $h'_2$. 
Imposing various symmetry restrictions, the interference fragmentation 
can be cast into a double density matrix~\cite{jjt1}
\begin{eqnarray}
{d^2 \hat{\cal M}\over dz\, dm^2}
= &&
\Delta_0(m^2)\left\{I\otimes \bar\eta_0\,
\hat{q}_I(z,m^2)
+\left(\sigma_+\otimes \bar\eta_-
+ \sigma_-\otimes \bar\eta_+\right)\delta\hat{q}_I(z,m^2)\right\}
\Delta^*_1(m^2)
\nonumber
\\
& &
+\Delta_1(m^2)\left\{I\otimes \eta_0\,
\hat{q}_I(z,m^2)
+\left(\sigma_-\otimes \eta_+ 
+ \sigma_+\otimes\eta_-\right)\delta\hat{q}_I(z,m^2)
\right\}\Delta^*_0(m^2)\ ,
\label{fragmentation}
\end{eqnarray}
where $\sigma_\pm\equiv (\sigma_1\pm i\sigma_2)/2$ with
$\{\sigma_i\}$ the usual Pauli matrices. The $\eta$'s are $4\times 4$ 
matrices in $(\sigma, \rho)$  helicity space with nonzero elements only 
in the first column,
and the $\bar\eta$'s are the transpose matrices 
 ($\bar\eta_0 = \eta_0^T, \bar\eta_+=\eta_-^T,
\bar\eta_-=\eta_+^T$), with the first rows $(0,0,1,0)$,
$(0,0,0,1)$, and $(0,1,0,0)$ for $\bar\eta_0$, $\bar\eta_+$, and $\bar\eta_-$,
respectively. The final state interactions between the two pions are included
explicitly in $\Delta_0(m^2)=-i \sin\delta_0 e^{i\delta_0}$ and 
$\Delta_1(m^2)=-i \sin\delta_1 e^{i\delta_1}$, with $\delta_0$ and $\delta_1$ 
the strong interaction $\pi\pi$ phase shifts. Here we have suppressed the 
$m^2$ dependence of the phase shifts.
Note that $\hat q_I$ and 
$\delta \hat q_I$ depend on both $m^2$ and $z$.  We expect that the 
principal $m^2$ dependence will enter through the final state factors 
$\Delta_0$ and $\Delta_1$.  To preserve
clarity, the $Q^2$ dependence  of the fragmentation functions 
has been suppressed.  Henceforth we suppress the $m^2$ dependence 
as well.

The decay density matrix, quark distribution function, and hard 
scattering cross section have been given explicitly in Ref.~\cite{jjt1}. 
For completeness we quote the results relevant to the present analysis. 
The interference part of the decay density matrix (after
integration over the azimuthal angle $\phi$) is 
\begin{equation}
{d {\cal D}\over d\cos\Theta}
=-{\sqrt{3}\over 2\pi m}
\, \cos\Theta \left(\bar\eta_0 +\eta_0\right)\ .
\end{equation}
The quark distribution function ${\cal F}$ is expressed as \cite{jaffe96}
\begin{equation}
{\cal F} = {1\over 2} q(x)~I\otimes I + {1\over 2} \Delta
q(x)~\sigma_3 \otimes 
\sigma_3+{1\over 2}  \delta q(x)~
\left(\sigma_+\otimes\sigma_-+\sigma_-\otimes\sigma_+\right)\ ,
\label{calf}
\end{equation}
where the first matrix in the direct product is in the
nucleon helicity space and the second in the quark helicity space.
Here $q(x)$, $\Delta q(x)$, and $\delta q(x)$ are the spin average, 
spin, and transversity distribution functions, 
respectively, and their dependence on $Q^2$ has been suppressed. Finally,
for a longitudinally polarized electron beam,
the hard scattering cross section is given by~\cite{jaffe96}
\begin{eqnarray}
{d^2\sigma_\pm (e q\rightarrow e' q')\over dx\, dy}&=& 
\frac{e^4e_q^2}{8\pi Q^2}\Biggl[\frac{1+(1-y)^2}{2y}
\left(I\otimes I + \sigma_3 \otimes 
\sigma_3\right)  
+  { 2(1-y)\over y}
\left(\sigma_+\otimes\sigma_-+\sigma_-\otimes\sigma_+\right)
\nonumber
\\*[7.2pt]
& &\hspace*{1cm}\pm { 2-y\over 2} \left(\sigma_3 \otimes I
+I\otimes \sigma_3\right)\Biggr]\ ,
\label{sigmahel}
\end{eqnarray}
in the collinear approximation, where the $\pm$ sign 
refers to the initial electron helicity.  Here $e_q$ is the charge fraction 
carried by a quark, and we have integrated out the azimuthal angle of 
the scattering plane.

The cross section can be obtained by putting all the
ingredients together. To facilitate our discussions, we
define forward and backward cross sections, $d\sigma^{\rm F}$ and 
$d\sigma^{\rm B}$, where the $\Theta$ dependence has been integrated 
over the forward ($0\leq\Theta\leq\pi/2$) and  backward 
($\pi/2\leq\Theta\leq\pi$) hemisphere in the two-pion rest frame,
respectively.
For a longitudinally polarized nucleon target with a longitudinally 
polarized electron beam, we then obtain the following double asymmetry
\begin{eqnarray}
{\cal A}^{\rm FB}_{\Uparrow\Downarrow} \equiv 
{\left(d\sigma^{\rm F}_{\uparrow\Uparrow} 
-d\sigma^{\rm B}_{\uparrow\Uparrow}\right)
-\left(d\sigma^{\rm F}_{\uparrow\Downarrow} 
-d\sigma^{\rm B}_{\uparrow\Downarrow}\right)
\over \left(d\sigma^{\rm F}_{\uparrow\Uparrow} 
+d\sigma^{\rm B}_{\uparrow\Uparrow}\right)
+\left(d\sigma^{\rm F}_{\uparrow\Downarrow} 
+d\sigma^{\rm B}_{\uparrow\Downarrow}\right)}
&=& -{\sqrt{3}\, y (2-y)\over 1+(1-y)^2}\,\,
\sin{\delta_0} \sin{\delta_1}
\cos\left(\delta_0-\delta_1\right)\, 
\nonumber
\\*[7.2pt]
& &\times
{\sum_a e_a^2 \Delta q_a(x)\, \hat{q}_I^a(z)\over
\sum_a e^2_a q_a(x)
\left[ \sin^2\delta_0
\hat{q}_0^a(z)
+\sin^2\delta_1
\hat{q}_1^a(z)\right]}\ ,
\label{asymmetry}
\end{eqnarray}
where $\hat{q}_0$ and $\hat{q}_1$ are spin-average fragmentation functions
for the $\sigma$ and $\rho$ resonances, respectively. The arrows
$\Uparrow$ and $\Downarrow$ ($\uparrow$ and $\downarrow$) indicate the 
nucleon's (electron's) polarization along and opposite to the beam direction.
Note that in Eq.~(\ref{asymmetry}) the electron's polarization is fixed
and the asymmetry requires flipping the nucleon's polarization. 
This result also holds when the electron and nucleon
polarizations are exchanged.

\begin{figure}[h]
\begin{minipage}[h]{6.0in}
\epsfxsize=7.0truecm
\centerline{\epsffile{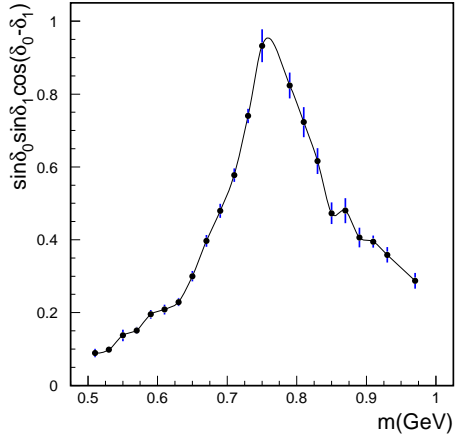}}
        \caption{{\sf The factor, $\sin\delta_0 \sin\delta_1
\cos(\delta_0-\delta_1)$, 
           as a function of the invariant mass $m$ of two-pion
system. 
           The data on $\pi\pi$ phase shifts are taken from
Ref.~\protect\cite{martin74}.}}
\label{fig2}
\end{minipage}
\end{figure}

The flavor structure of the asymmetry 
${\cal A}^{\rm FB}_{\Uparrow\Downarrow}$ is quite simple.  
Isospin symmetry gives $\hat{u}_I =
- \hat{d}_I$ and $\hat{s}_I = 0$, and charge
conjugation requires $\hat{q}^a_I=- \hat{\bar
q}^a_I$. This implies $\sum_a\, e_a^2\,\Delta
q_a\,\hat{q}_I^a=[{4/ 9}\, \Delta u_v- {1/9}\, \Delta
d_v]\,\hat{u}_I$, where $\Delta q_v \equiv \Delta q -
\Delta\bar q$. 
Therefore ${\cal A}^{\rm FB}_{\Uparrow\Downarrow}$ is sensitive
to the valence quark spin distribution. Note that the interference 
between the two partial waves makes the interference fragmentation
function $\hat{q}_I^a(z)$ charge conjugation odd
and accesses the valence quark spin distribution.
The role of final state interactions is quite different here than in
Ref.~\cite{jjt1}.  Here the effect persists as long as two partial waves 
of opposite C-parity are active, whether or not they are out of phase.  
This is evidenced in Eq.~(\ref{asymmetry}) by the 
factor $\cos(\delta_0-\delta_1)$.  
Note that if either $\delta_0$ or $\delta_1$ goes to 
zero and thus only one partial wave is active, the asymmetry vanishes 
as required by charge conjugation.  
In Fig.~\ref{fig2} we have plotted
the factor, $\sin\delta_0\sin\delta_1\cos(\delta_0-\delta_1)$, as a
function of the two-pion invariant mass $m$. It is positive over the
$\rho$ region and hence the effect does not average to
zero over this region. This differs from the case of
the transversity asymmetry derived in Ref.~\cite{jjt1}, where the
interference averages to zero over the $\rho$ region due to a
factor $\sin(\delta_0-\delta_1)$. We also see from Fig.~\ref{fig2}
that the interference peaks near the $\rho$ mass, indicating
that an optimal signal would be in the vicinity of $m\sim m_\rho$. It
is unclear at this stage whether the effect would survive averaging
over the $z$ dependence of the interference fragmentation function.

A similar forward-backward asymmetry appears in the unpolarized 
process
\begin{eqnarray}
{\cal A}^{\rm FB} 
\equiv {d\sigma^{\rm F}-d\sigma^{\rm B}
\over d\sigma^{\rm F}+d\sigma^{\rm B}}
&=& -\sqrt{3} \,\,
\sin{\delta_0} \sin{\delta_1}
\cos\left(\delta_0-\delta_1\right)\,
\nonumber
\\*[7.2pt]
& &\times
{\sum_a e_a^2  q_a(x)\, \hat{q}_I^a(z)\over
\sum_a e^2_a q_a(x)
\left[ \sin^2\delta_0
\hat{q}_0^a(z)
+\sin^2\delta_1
\hat{q}_1^a(z)\right]}\ .
\label{asymmetry1}
\end{eqnarray}
Isospin symmetry and charge conjugation again dictate that
$\sum_a\, e_a^2\,q_a\,\hat{q}_I^a=[{4/ 9}\,   u_v- 
{1/9}\,  d_v]\,\hat{u}_I$ for $\pi^+\pi^-$ system,
where $ q_v \equiv q -  \bar q$. Given that the valence quark distributions
$u_v$ and $d_v$ are known experimentally, this asymmetry
can be used to determine the interference fragmentation 
function $\hat{q}_I^a(z)$ and hence isolate the
valence quark spin distribution from ${\cal A}^{\rm FB}_{\Uparrow\Downarrow}$.
This point becomes more transparent when 
${\cal A}^{\rm FB}_{\Uparrow\Downarrow}$ is combined with
${\cal A}^{\rm FB}$. 
In this case,  one can find an asymmetry {\it independent} 
of the interference fragmentation functions:
\begin{eqnarray}
{\left(d\sigma^{\rm F}_{\uparrow\Uparrow} 
-d\sigma^{\rm B}_{\uparrow\Uparrow}\right)
-\left(d\sigma^{\rm F}_{\uparrow\Downarrow} 
-d\sigma^{\rm B}_{\uparrow\Downarrow}\right)
\over 
\left(d\sigma^{\rm F}_{\uparrow\Uparrow} 
-d\sigma^{\rm B}_{\uparrow\Uparrow}\right)
+\left(d\sigma^{\rm F}_{\uparrow\Downarrow} 
-d\sigma^{\rm B}_{\uparrow\Downarrow}\right)}
&=& {y(2-y)\over 1+(1-y)^2}\,\,\,
{4 \Delta u_v - \Delta d_v\over 4 u_v - d_v}\ .
\label{asymmetry2-pi}
\end{eqnarray}
So, one may use $\pi^+\pi^-$ production on both
nucleon and deuteron targets to measure $\Delta u_v$ and $\Delta d_v$. 
Although the results in Eq.~(\ref{asymmetry2-pi})
are independent of $z$, one should 
keep the cross section differential in $z$ to avoid
possibly washing out the effect.

Refs. \cite{frankfurt89,close91} explain how
to use single meson (pion or kaon) production in deep inelastic
scattering to measure the valence quark spin distribution in
the nucleon.   It is clear from Eq.~(\ref{asymmetry2-pi}) 
that our asymmetry is a  particular contribution to the ones described 
in Refs.~\cite{frankfurt89,close91}.  Ours is perhaps more under control 
since it is differential in $m^2$ and expressed in terms of $\pi\pi$ phase 
shifts.  The single particle asymmetry makes use of a larger data set.  These 
independent ways of
measuring the valence quark spin distributions should both be pursued.
In addition, the asymmetries we have studied are sensitive to two particle 
interference fragmentation functions which may be interesting quantities in 
their own right.
These measurements may be carried out
in facilities such as HERMES at HERA and COMPASS at CERN, both of which
have sensitivity to the hadronic final state in electron scattering.

To summarize,  we have discussed the applications of the twist-two 
interference quark fragmentation functions introduced previously
to the case of longitudinally polarized electron beam and 
longitudinally polarized nucleon target. We obtain two asymmetries:
one provides a probe of the valence quark spin distribution, and the 
other can be used to extract the interference fragmentation functions.


\vspace*{1cm}

We would like to thank John Collins for pointing out an error 
in the earlier version of this paper.

\end{document}